\documentclass{blois}

\usepackage{amsmath}

\def\be{\begin{equation}}
\def\ee{\end{equation}}
\def\bea{\begin{eqnarray}}
\def\eea{\end{eqnarray}}

\newcommand{\C}{\mathcal{C}}

\newcommand{\pd}{\partial}

\newcommand{\mc}{\mathcal}
\newcommand{\tx}{\mathrm}

\newcommand{\td}{\text{d}}
\newcommand{\vp}{\varphi}
\newcommand\Q{{\cal Q}}

\newcommand{\Photo}{}

\begin{document}
\vspace*{4cm}
\title{Compact objects in gravity theories}

\author{Athanasios Bakopoulos$^{2}$, Christos Charmousis$^{1}$ and  Nicolas Lecoeur$^{1}$}

\address{$^{1}$ Universit\'e Paris-Saclay, CNRS/IN2P3, IJCLab, 91405 Orsay, France \\
    $^{2}$ Division of Applied Analysis, Department of Mathematics,
University of Patras, Rio Patras GR-26504, Greece}

\maketitle
\abstracts{We briefly discuss explicit compact object solutions in higher-order scalar-tensor theories. We start by so-called stealth solutions, whose metric are General Relativity (GR) solutions, but accompanied by a non-trivial scalar field, in both spherically-symmetric and rotating cases. The latter then enables to construct an analytic stationary solution of scalar tensor theory which is called disformed Kerr metric. This solution constitutes a measurable departure from the usual Kerr geometry of GR. We finally consider a scalar-tensor theory stemming from a Kaluza-Klein reduction of a higher-dimensional Lovelock theory, and which enables to obtain non-stealth black holes, highly compact neutron stars and finally wormhole solutions.}

\section{Introduction}

There is an impressive income of observational data for compact objects, namely neutron stars and black holes. These observations have initiated a revolutionary epoch in the field of gravitational physics. Indeed, since a few years, we have been observing gravitational waves (GW) emitted from binary mergers (see in particular \cite{LIGOScientific:2017vwq}, \cite{LIGOScientific:2020zkf}, \cite{LIGOScientific:2021qlt}).  They concern compact objects of relatively few solar masses. Moreover, we have now images of supermassive black holes generated from networks of radio-telescopes \cite{EventHorizonTelescope:2020qrl} such as the Event Horizon Telescope (EHT). As regards X-ray observations, like the NICER mission (Neutron star Interior Composition ExploreR), they strive to gather information on the equation of state (EoS) of neutron stars by observing their thermal hotspots. Furthermore, GRAVITY, operating in Chile since 2016, is a very large telescope interferometer, combining light from four different telescopes. Anchoring its position to a very bright star, it can follow stars in its vicinity with better and better precision for long exposure times. In this way, GRAVITY gathers information on stars, such as S2, orbiting our Galactic center \cite{GRAVITY:2018ofz}. Such or similar observational results are quite compatible with predictions emanating from General Relativity (GR). Even at this early stage however, certain questions do arise, regarding for instance the nature of the secondary object in the binary of the signal GW190814 \cite{LIGOScientific:2020zkf}: its mass of $2.59_{-0.09}^{+0.08}~M_\odot$ places it in the current observational mass gap predicted by GR, in between neutron stars and black holes. Such a compact object could be explained (within GR) only as a neutron star with an unexpectedly stiff (or exotic) EoS, disfavoured by \cite{LIGOScientific:2017vwq}, a neutron star with an unexpectedly rapid rotation, or a black hole with an unexpectedly small mass (for a recent discussion, see \cite{Charmousis:2021npl} and references within). Future observational data, evolving from discovery towards precision, will  permit to probe additional gravitational parameters, eventually checking the validity of no-hair theorems, star trajectories beyond precession, novel observations of exceptional objects, including binary pulsars in a strong gravity field, etc. One could possibly entertain the discovery of novel compact objects theoretically disfavoured from GR such as wormholes (for example distinguishing a wormhole throat versus an event horizon, see e.g. \cite{Cardoso:2016rao}).

\section{Black holes in GR}\label{subsec:prod}

In GR, black holes are "unique", and characterized by a finite number of charges, namely the mass $M$ and the angular momentum $J$, putting aside electric and magnetic charge (EM). During collapse, black holes lose their hair and relax to some stationary state of large symmetry. Therefore, omitting EM charge, stationary black holes are vacuum solutions of Einstein's equations. They are relatively simple solutions -they have no hair- and the static and spherically symmetric black hole is uniquely determined by its mass $M$ as the unique Schwarzschild solution,  $$\td s^2=-f(r)\td t^2+\frac{\td r^2}{f(r)}+ r^2 \td \Omega^2,$$ where $f(r)=1- \frac{2M}{ r}$. The zero of $f(r)$ is the event horizon of the black hole ($r_h=2 M$), determining the absolute surface of no return for test particles or light. The interior of the event horizon is the trapped region of the black hole, hiding  the curvature singularity at $r=0$ from the outside observers. The photon sphere residing at $r_p= \frac{3}{2} r_h$ is determined by light geodesics giving the celebrated light ring of the Schwarschild black hole, compatible with the observation of M87  by the EHT a few years ago. 

Once we allow for rotation, the angular momentum $J$ and mass $M$ are the sole parameters of the unique Kerr black hole \cite{Kerr:1963ud} whose metric reads,
\begin{eqnarray*}
\label{kerr}
\td s^2 = -\left(1-\frac{2Mr}{\rho^2}\right)\td t^2 -\frac{4aMr \sin^2\theta}{\rho^2} \td t \td \vp + \frac{\sin^2\theta}{\rho^2}\left[(r^2+a^2)^2 -a^2 \Delta \sin^2\theta\right]\td \vp^2\\
+ \frac{\rho^2}{\Delta}\td r^2 + \rho^2 \td\theta^2,
\end{eqnarray*}
where $\rho^2 = r^2 + a^2\cos^2\theta,\;
\Delta = r^2 + a^2 - 2Mr\; .$

"No hair" in practise means that any additional parameter describing fluctuations of the geometry is given by a combination of these two parameters. As such, for a Kerr spacetime, the quadrupole is fixed in a unique way as $Q^2=-J^2/M$. Therefore, any simultaneous measure of these three quantities is a direct check for the validity of Kerr spacetime. 

The Kerr geometry has a number of interesting properties, we will here briefly refer to the "integrability" of its geodesics, as it will be important for following considerations. This property was demonstrated in a celebrated paper by Brandon Carter back in 1968 \cite{Carter:1968rr}. But what is the precise meaning of integrability  here? Using the Hamilton-Jacobi (HJ) formalism we may, symmetries permitting, write the geodesic equations which are a priori second order, as a system of first order equations with respect to an affine parameter. For this integrability property to be true, one needs to have 4 constants of motion (for 4 dimensions). For Kerr, there are 3 obvious constants of motion: $E$, $L_z$, $m$, energy at infinity, angular momentum in the rotation axis and rest mass of the test particle. So in principle, one of the dimensions needs to be fixed, in order to study geodesics in a HJ fashion, by restricting for example to motion in the Kerr equator.

The HJ equation reads,
\be
\label{HJ}
\frac{\partial S}{\partial \lambda} =   g^{\mu\nu} \frac{\partial S}{\partial x^\mu}   \frac{\partial S}{\partial x^\nu} =-m^2,
\ee
where $\lambda$ is the affine parameter of the curve, and the HJ functional $S$ is defined as,
\be
\frac{\partial S}{\partial x^\mu}=p_\mu,
\ee
with $p^\mu$ the four-momentum of the test particle. Since $p_t=-E$ and $p_\varphi=L_z$, the HJ functional is given by
\be
\label{carter}
S(t,\varphi,r,\theta)=-E t+ L_z \varphi+ S(r,\theta).
\ee
Brandon Carter showed \cite{Carter:1968rr} that $S(r,\theta)=S_r (r) + S_\theta (\theta)$ is actually separable, and the fourth constant of motion, denoted by $\Q$, is the Carter's separation constant of motion. Hence we have integrability and $S$ is a known function for Kerr. Its precise form will lead us to a rotating hairy solution for scalar-tensor theories as we shall see.

\section{Horndeski theory}

We will be discussing scalar-tensor theories for the remaining part of this paper. Scalar-tensor theories are probably the most robust and simple extension of GR, as they have a unique additional degree of freedom, a real scalar and a quite general action. We regard them not as fundamental, in fact their UV properties are not much better than GR, but they are well defined theories and consist a measurable departure from GR. Furthermore, most UV modified gravity theories of more fundamental nature, e.g. originating from string theory, extra dimensions, bigravity/massive gravity, acquire some limit where they become scalar-tensor theories. As such, one can expect some similarities in phenomenology to be present in scalar-tensor theories. An additional advantage is their mathematical robustness. Indeed the most general scalar-tensor theory with second order equations of motion was found a long time ago by Horndeski as a Lovelock type theorem for 4 dimensions \cite{Horndeski:1974wa}. Horndeski theory is parametrized by four functions $\{G_i: i=2,..,5\}=\{G_2,G_3,G_4,G_5\}$ of the scalar $\phi$ and its kinetic density $X=-\frac{1}{2}\partial_\mu\phi\,\partial^\mu\phi=-\frac{1}{2}\nabla_\mu\phi\,\nabla^\mu\phi$,
\be
\label{eq:Hfr}
S_\tx{H} = \displaystyle\int \tx{d}^4x \sqrt{-g} \,\left(\mc{L}_2+\mc{L}_3+\mc{L}_4+\mc{L}_5 \right),
\ee
with
\bea
&\mc{L}_2 =G_2(\phi,X) ,
\label{eq:L2fr}
\\
&\mc{L}_3 =-G_3(\phi,X) \,\nabla^\alpha\nabla_\alpha \phi ,
\label{eq:L3fr}
\\
&\mc{L}_4 = G_4(\phi,X) R + G_{4X} \left[ (\nabla^\alpha\nabla_\alpha \phi)^2 -\nabla_\mu\nabla_\nu\phi \,\nabla^\mu\nabla^\nu\phi\right] ,
\label{eq:L4fr}
\\
&\mc{L}_5 = G_5(\phi,X) G_{\mu\nu}\nabla^\mu \nabla^\nu \phi - \frac{1}{6}\, G_{5X} \big[ (\nabla^\alpha\nabla_\alpha \phi)^3 - 3\,\nabla^\alpha\nabla_\alpha \phi\, \nabla_\mu\nabla_\nu\phi\,\nabla^\mu\nabla^\nu\phi
\\
&\quad + 2\,\nabla_\mu\nabla_\nu\phi\, \nabla^\nu\nabla^\rho\phi\, \nabla_\rho\nabla^\mu\phi \big].
\label{eq:L5fr}
\eea

As stressed above, this action is just a parameterization of the most general scalar-tensor theory, and the theory is defined up to integrations by parts.  The equations of motion are however unique. Note that $G_2$ and $G_4$ have parity symmetry in $\phi$, unlike $G_3$ and $G_5$. It turns out that compact objects within $G_2$ and $G_4$ are more closely related to GR, whereas $G_3$ and $G_5$ have more distinct phenomenology (see for example \cite{Doneva:2017bvd}) and are far more difficult to tackle analytically \cite{Bakopoulos:2022csr}. We will gain some understanding of this further on in our review. Note also that when the functions $G_i$ only depend on $X$, the theory has shift symmetry in $\phi$. Before pushing on further, let us insist on the importance of conformal and disformal transformations. In fact conformal and disformal transformations transport us between theories in an unexpected and interesting way \cite{Zumalacarregui:2013pma}. 

Consider the mapping 
\be
\label{CD}
g_{\mu\nu}\longrightarrow \tilde{g}_{\mu\nu}=C(\phi, X) g_{\mu\nu}+ D(\phi, X) \nabla_\mu \phi\, \nabla_\nu \phi, 
\ee
 for given (regular) functions $C$ (conformal) and $D$ (disformal) of $\phi$ and $X$. One can show that a conformal and disformal mapping depending only on $\phi$ is an internal Horndeski map. However, a disformal and conformal map depending on $X$ venture us out of Horndeski, in so-called beyond Horndeski \cite{Gleyzes:2014dya} and DHOST theories \cite{Langlois:2015cwa} respectively. In other words, there are more general theories than Horndeski's theory, which still are well defined scalar-tensor theories and display a single metric and scalar as sole gravitational degrees of freedom: in spite of higher-order equations of motion, they have been shown to propagate only one additional, scalar degree of freedom. They are related to Horndeski by the above map (\ref{CD}).

\section{Shift and parity symmetric theories}

We shall first consider an example theory that has shift and parity symmetry for the scalar field $\phi$,
\begin{equation}
\label{john}
S = \int \td^4x \sqrt{-g}\left[R  - 2 \Lambda_b -\eta X +\beta\, G^{\mu\nu}\nabla_\mu\phi\, \nabla_\nu\phi  \right].
\end{equation}
This theory has linear $G_2$ and $G_4$ terms when written in the form (\ref{eq:Hfr}).
One can look for the general static and spherically symmetric spacetime solution, 
\be
\td s^2=-h(r)\td t^2+\frac{\td r^2}{f(r)}+ r^2 \td \Omega^2,
\ee
allowing however a time and radially dependent scalar field $\phi=\phi(t,r)$. The general solutions for this symmetry are found and shown to be classified as solutions of a third order algebraic equation \cite{Babichev:2013cya}. 
 A particular solution acquires the following {\it{stealth}} GR form,
$$f  =h = 1- \frac{2\mu}{r} + \frac{\eta}{3\beta}r^2,$$ with a linearly time dependent scalar field, $$\phi=q\,t \pm \int \td r \; \frac{q}{h}\sqrt{1-h}\, .$$ {\it{Stealth}} signifies that the metric has an identical form to the standard de Sitter-Schwarzschild geometry of GR, while the scalar is non trivial. Note that the secondary scalar hair parameter, $q$, is fixed by $q^2=\frac{\zeta \eta+\Lambda_b \beta}{\beta \eta}$. Here, it plays the role of the self tuning integration constant relating the bare $\Lambda_b$ with the coupling constants of the theory, $\eta$ and $\beta$ \cite{Charmousis:2015aya}. It is important to note that the kinetic term, $-2X=g^{\mu \nu} \pd_\mu\phi\,\pd_\nu \phi=-\frac{q^2}{h} +q^2\frac{f(1-h)}{h^2}=-q^2$, is a constant for the stealth solutions, and this permits to generalize the solution to arbitrary parity and shift symmetric theories \cite{Kobayashi:2014eva}. One can show quite easily that a disformed version of the above solution is again a stealth black hole \cite{Babichev:2017lmw}. Hence, stealth solutions of spherical symmetry will be generic in shift and parity symmetric DHOST theories, while not giving rise to novel black hole geometries. 

What about rotating solutions? Can we find a scalar tensor theory where a stealth Kerr black hole exists? In other words is there a theory admitting Kerr as a metric solution with a non trivial scalar field? It turns out that the non linear leap from staticity to stationarity is generically very difficult. For our sample Horndeski theory (\ref{john}), this is impossible beyond slow rotation. It is only by venturing to DHOST that such a theory exists. The theory in question is the subclass of DHOST where gravitational waves propagate at the speed of light \cite{Langlois:2015cwa}. One can show that under minor requirements, such a theory can acquire a Carter \cite{Carter:1968ks} (de Sitter-Kerr) or Kerr \cite{Kerr:1963ud} stealth solution, with $X$ constant \cite{Charmousis:2019vnf}. The real difficulty is finding a regular scalar field with $X$ constant for a Kerr or Carter geometry. This problem is in fact related to the integrability of the Kerr geodesics. Simply note that the identification $\phi=S$, where $S$ is the separable HJ functional of Eq. (\ref{carter}), constitutes the most general scalar field parametrized by 4 constants of integration such that $X$ is constant, pursuant to the HJ equation (\ref{HJ}). The regular scalar field for Kerr \cite{Kerr:1963ud} and Carter metrics of positive effective cosmological constant \cite{Carter:1968ks} was found in \cite{Charmousis:2019vnf}, where the interested reader can find details of the construction. For the case of the Kerr stealth metric, the scalar field, regular at the event horizon (in Kerr coordinates) reads,
\be
\label{S}
\phi(t,r) = q\, t +  \int \td r\,\frac{\sqrt{q^2 (r^2+a^2)2 M r}}{\Delta}  \,, 
\ee
where we have set $E=m=q$, $L_z=0$, $\Q=0$ for the "geodesic" trajectory parameters in order to achieve regularity. By switching off rotation, $a=0$, we get back to the spherically symmetric scalar field. 
It is interesting to note that, while a positive cosmological constant allows for a two parameter family of Carter scalars, the Kerr stealth acquires only one, $q$ (as for spherical symmetry), whereas for a negative cosmological constant a stealth solution is in fact impossible as there is no HJ functional regular throughout the static black hole region.

As opposed to spherical symmetry, a disformal transformation of the stationary stealth Kerr is not trivial, leading to a different stationary geometry with the same scalar field (\ref{S}). In fact, the disformed Kerr metrics \cite{Anson:2020trg} are
$$g^{Kerr}_{\mu\nu}\longrightarrow \tilde{g}_{\mu\nu}=g^{Kerr}_{\mu\nu}+D(X) \nabla_\mu \phi\, \nabla_\nu \phi,$$ where the disformal mapping is in fact constant because $X$ itself is constant, $D\left(X\right)\equiv D$ and the scalar field remains the same from seed to target solution (\ref{S}). Each constant $D$ corresponds to a different DHOST theory, but we can choose to treat $D$ as a deformation parameter of Kerr geometry. 
Disformed Kerr is a stationary solution which, under certain requirements, is a black hole with good causality properties. In other words, it has similar characteristics to the Kerr solution, while acquiring a number of distinct features. The metric reads,
\bea
\td s^2= -\left(1-\frac{2\tilde{M}r}{\rho^2}\right)\td t^2 -\frac{4\sqrt{1+D}\tilde{M}ar\sin^2\theta}{\rho^2}\td t\td\vp + \frac{\sin^2\theta}{\rho^2}\left[\left(r^2+a^2\right)^2-a^2\Delta
\sin^2\theta\right]\td\vp^2 \nonumber \\
 +  \frac{\rho^2 \Delta - 2 \tilde{M}(1+D) r D (a^2+r^2)}{\Delta^2}\td r^2 - 2D \frac{\sqrt{2\tilde{M}r(a^2+r^2)}}{\Delta}\td t\td r + \rho^2 \td \theta^2\; ,
 \eea
where $\tilde{M}=M/\left(1+D\right)$. For a start, when $D\neq 0$ and $a\neq 0$, the disformed metric is not an Einstein metric! Furthermore, when $D\neq 0$, the Kerr no hair relation is no more verified \cite{Anson:2021yli}, and each $D$ leads to a distinct quadrupole moment $Q$. The disformed metric is not circular, and its geodesics are not integrable. It constitutes therefore a measurable departure from Kerr geometry which for the moment agrees with observational data but may yield observable features in the future.  As a last remark we note that within DHOST theories one can construct regular black hole geometries  upgrading  Kerr-Schild techniques \cite{Babichev:2020qpr}.

\section{Adding in parity breaking terms}

So much for parity and shift symmetric theories including the $G_2$ and $G_4$ terms of Horndeski. For parity breaking theories, including $G_3$ and $G_5$ terms, finding analytic solutions seems quite difficult. Their interest is no lesser however, as they include theories with interesting phenomenology as the Gauss-Bonnet term. One way of analytic approach is through a Kaluza-Klein reduction of higher-dimensional Lovelock theory. One can then obtain analytic solutions emanating from higher-dimensional Lovelock black holes \cite{Charmousis:2012dw}. They are however characterized by a non-Newtonian mass fall-off, which is to be expected since they originate from higher-dimensional solutions. The way out of this phenomenological restriction problem came from the work of Glavan and Lin \cite{Glavan:2019inb}, who realized that a non-trivial singular limit may be taken, yielding new, non-trivial metrics in 4 dimensions (where Lovelock theory is exactly GR). The relevant reduced scalar-tensor theory was then constructed by Lu and Pang \cite{Lu:2020iav} and later generalized by Fernandes \cite{Fernandes:2021dsb} who constructed the most general scalar-tensor theory where the scalar field is conformally coupled. Here we will focus on the former theory which reads,
\begin{equation}
S=\displaystyle\int{\mathrm{d}^4x\sqrt{-g}\left\{R+\alpha\left[\phi\mathcal{G}+4G_{\mu\nu}\nabla^\mu\phi\nabla^\nu\phi-4(\nabla\phi)^2\nabla^\alpha\nabla_\alpha \phi+2(\nabla\phi)^4\right]\right\}}+S_\mathrm{m},\nonumber
\end{equation}
and harbors an overall extra coupling constant $\alpha$ of dimensions of length squared. $\mathcal{G}$ is the Gauss-Bonnet invariant, $\mathcal{G}=R^2-4R_{\mu\nu}R^{\mu\nu}+R_{\mu\nu\rho\sigma}R^{\mu\nu\rho\sigma}$. 
For static and spherical symmetry, Lu and Pang found the following black hole solution, 
\be
\label{LP}
\td s^2=-f(r)\td t^2+\frac{\td r^2}{f(r)}+r^2\td \Omega^2, \quad f(r)=1+\frac{r^2}{2 \alpha }\left(1-\sqrt{1+ \frac{8 \alpha  M} {r^3}}\:\right),
\ee 
while
$\phi(r)=\displaystyle\int{\mathrm{d}r\,\dfrac{\sqrt{f}- 1}{r\sqrt{f}}}$. The solution can be extended for the scalar to a linear time dependence without a change in the spacetime metric $f$ \cite{Charmousis:2021npl}. This renders the scalar well defined within and on the horizon \cite{Charmousis:2021npl} where $f\leq 0$. The solution is most importantly not a stealth solution, and a direct calulation shows that $X$ is no longer constant as long as $M\neq 0$. 
Far away as $r\to\infty$, the solution looks very much like GR, $f(r)=1-\frac{2 M}{r}+\frac{4 \alpha  M^2}{r^4}+\mathcal{O}(r^{-5})$ and as such has very similar weak gravity phenomenology to GR \cite{Clifton:2020xhc}. Its structure however is far more reminiscent of an electrically charged black hole (without charge), as it has an outer and inner horizons as long as $M>M_\mathrm{min}=\sqrt{\alpha}$. In addition, the solution is more regular than GR at $r\rightarrow 0$, although still not regular, as $f\left(r\right)\sim 1- \sqrt{\frac{Mr}{2\alpha}}$ (one would need a de Sitter core like behavior to have a regular black hole). In other words, we see that higher-order terms are partially smoothing out the geometry. In this direction, we will point out below that this theory allows greater compacity neutron stars! The coupling $\alpha<0$ is excluded from probed atomic nuclei which are horizonless. Indeed, nuclei radii $R\sim 10^{-15}\,\mathrm{m}$ implies a tiny $-\alpha < 10^{-30}\,\mathrm{m}^2$. For $\alpha>0$, since $M_\mathrm{min}=\sqrt{\alpha}$, then observed black holes from GW give us constraints on the magnitude of $\alpha$. For example, if the secondary object of GW190814 is a black hole, then $\alpha < 59\, \mathrm{km}^2$.

Introducing a perfect fluid interior, $T_{\mu\nu}=(\epsilon+P)u_\mu u_\nu+Pg_{\mu\nu}$, neutron star solutions can be found \cite{Charmousis:2021npl} which are everywhere regular. Quite nicely, for $\alpha>0$, this leads to more compact neutron stars than in GR. This fact simply alleviates tension from GW190814 if its secondary object is a neutron star, by allowing it to display a more simple and natural EoS than in the framework of GR, even with a slowly rotating neutron star. The most important and surprising result is the presence of a universal point of convergence for neutron stars and black holes, and this for {\it{generic EoS}}: in this theory, and as opposed to GR, there is no mass gap between neutron stars and black holes! This is a clear cut difference singling out this scalar tensor theory from GR with a single additional parameter $\alpha$. 
\begin{figure}
\centerline{\includegraphics[width=0.8\linewidth]{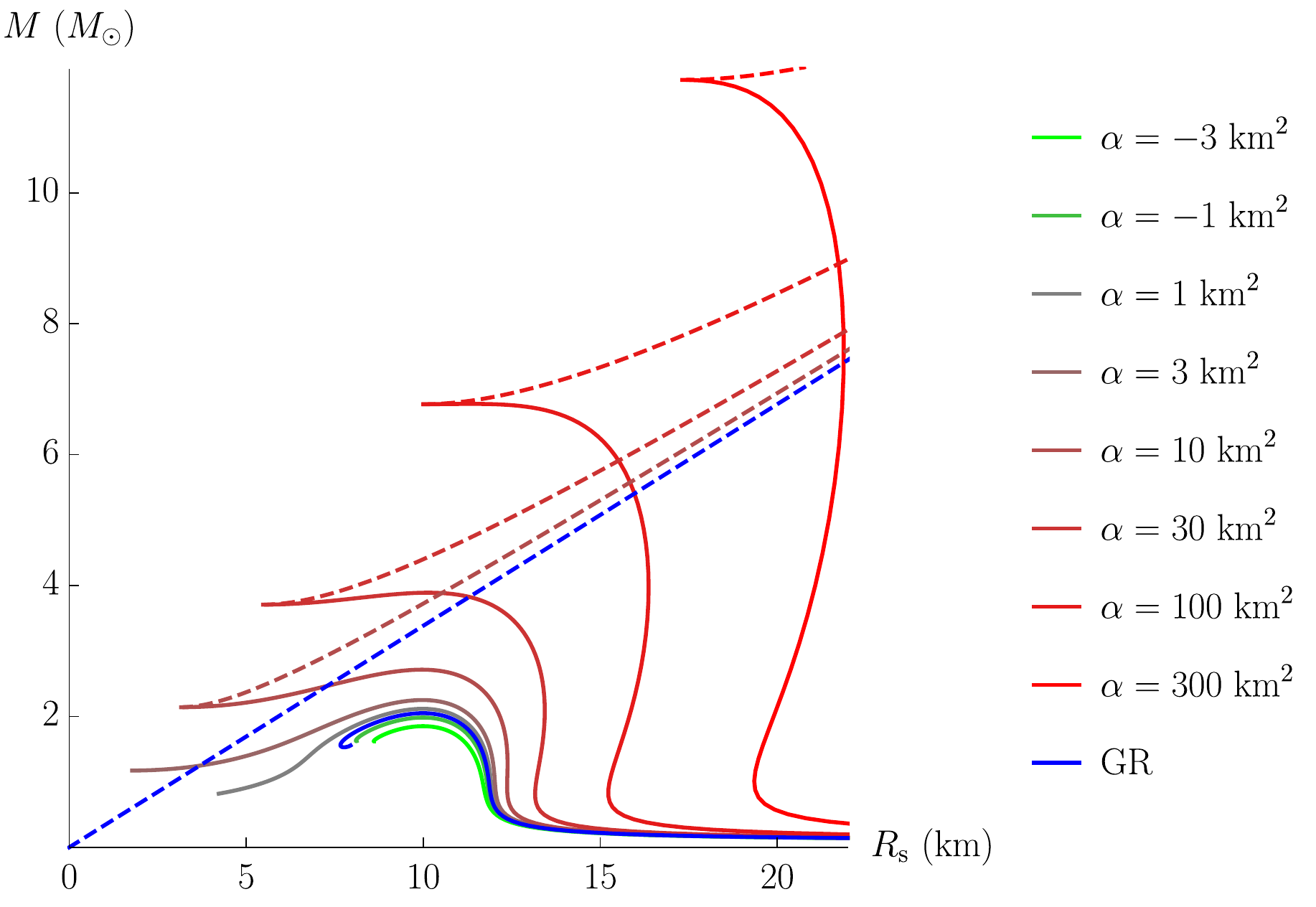}}
\caption[]{Mass-radius plot for the SLy EOS and various values of $\alpha$ taken from \cite{Charmousis:2021npl}. The plain blue line
are GR neutron star solutions while the dashed diagonal stands for the GR black hole. The other plain lines correspond to nonzero values of $\alpha$ while the dashed lines correspond to the location of black hole solutions. 
For  $\alpha>0$, the extreme compacity neutron star asymptotes the extremal black hole leaving no mass gap inbetween the compact objects in this theory unlike the case of GR.}
\label{fig:radish1}
\end{figure}
\begin{figure}
\centerline{\includegraphics[width=0.8\linewidth]{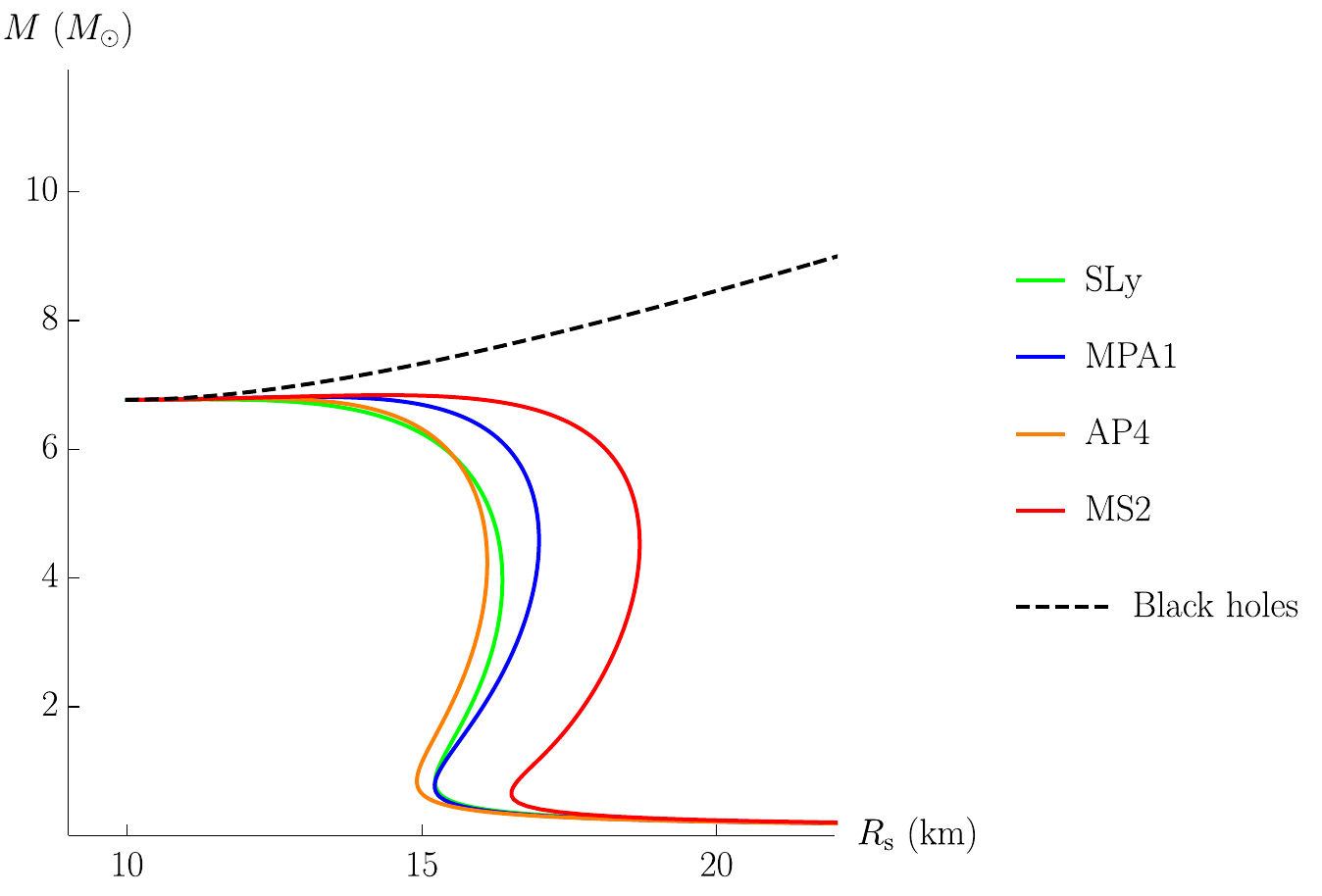}}
\caption[]{Mass-radius relations for fixed $\alpha = 100 \mathrm{km}^2$ and differing EOS taken from \cite{Charmousis:2021npl}. The black dashed line stands for black hole solutions, while coloured lines correspond to neutron star solutions. In spite of the variety
of behaviors at lower densities, the neutron star sequences associated to different EOS universally
converge towards the same endpoint, which is also the endpoint (extremal) of the black hole sequence.}
\label{fig:radish2}
\end{figure}

\section{Constructing wormholes}

Let us now move on to the construction of everywhere regular wormhole spacetimes. A wormhole has locally a similar spacetime geometry to a black hole. One can start with a spherically symmetric metric,
\be
\td s^2=-h(r)\td t^2+\frac{\td r^2}{f(r)}+r^2\td \Omega^2 ,
\ee
 but crucially, in order to have an inner and outer going throat, we need to have $f(r_T)=0$ while $h(r_T)> 0$ at the throat location $r=r_T$. The smaller $h(r_T)$ is, the greater the redshift and thus the closer the wormhole phenomenology is to a black hole, as the compact object essentially becomes a one way membrane. Indeed, for a wormhole, matter and light can in principle traverse the throat both ways. If the throat is (almost) an event horizon, $h\left(r_T\right)\sim 0$, then it becomes a one way wormhole. We can construct with relative ease such a geometry \cite{Bakopoulos:2021liw} by performing a disformal transformation starting with the black hole metric (\ref{LP}). 
Consider the map (\ref{CD}) with $C=1$ and $D=D(X)$, and seed metric  (\ref{LP}).
Indeed, if the scalar is only radially dependent while $X$ is not constant, a  disformal  transformation changes only the $g_{rr}$ term non trivially namely,  
\be
\tilde{h}=f\,, \quad
    \tilde{f}=\frac{f}{1+ 2D(X)  X} \equiv f \,W( X)^{-1}\,, \quad
      \tilde{\phi}=\phi \,.
      \ee
       In other words, we need $W(X)^{-1}=0|_{r=r_T}$ with $r_T> r_h$ (the seed event horizon location). 
       For example one can take \cite{Bakopoulos:2021liw},
\be
W(X)^{-1}=1-\frac{r_0}{\lambda}\,\sqrt{-2X}\,,
\ee
where $\lambda$ is a dimensionless quantity paramtrising how close our wormhole is to a black hole as, $h(r_0)=(1-\lambda)^2$.
If we can glue together two patches of $[r_T, \infty]$ in a $\C^2$-differentiable manner, then the obtained metric describes a regular traversable wormhole with a now, {\it{global}} timelike Killing vector. Choosing a generic shape function $W(X)$, wormholes of variable mass and redshift can indeed be constructed \cite{Bakopoulos:2021liw}.
The wormhole is found to be everywhere regular, and needs no local or non-local sources of matter. It is a vacuum solution of the obtained beyond Horndeski theory, just like the usual black holes are vacuum solutions of GR. The throat, unlike the event horizon, is always a light ring or critical point! Light will accumulate at the throat of the wormhole, and in many cases will have distinctive features such as multiple light rings (up to three).

\section*{Acknowledgments}
We thank T. Anson, E. Babichev, M. Crisostomi, R. Gregory, M. Hassaine,  P. Kanti,  A. Leh\'ebel, E. Smyrniotis and N. Stergioulas for discussions and collaboration on articles outlined here. CC thanks the organisers of the Rencontres de Blois for giving him the opportunity to present this work and for organising a very interesting conference. C.C. thanks the Laboratory of Astronomy of AUTh in Thessaloniki for hospitality during the course of this work.

\end{document}